\documentclass{aa}
  \usepackage{amsmath} 
  \usepackage[T1]{fontenc}
  \usepackage[varg]{txfonts}
  \usepackage{graphicx}
  \usepackage{natbib}
  \usepackage{array}
  \usepackage{xspace}

\usepackage[tight]{subfigure}

\def\BE{\begin{equation}}
\def\EE{\end{equation}}
\def\BA{\begin{array}}
\def\EA{\end{array}}
\def\BAN{\begin{eqnarray}}
\def\EAN{\end{eqnarray}}

\def\FIG #1 #2 [#3] #4\par{%
 \begin{figure}
   \resizebox{\hsize}{!}{\includegraphics[#3]{#2}}
   \caption{#4}
    \label{#1}
 \end{figure}
}

\def\FIGG #1 #2 #3 [#4] #5\par{%
 \begin{figure}[!h]
   \begin{center}
   \includegraphics*[#4]{#2}
   \includegraphics*[#4]{#3}
   \caption{\label{#1}#5}
   \end{center}
 \end{figure}
}

\def\FIGs #1 #2 #3 #4 #5 [#6] #7\par{%
 \begin{figure}[!h]
   \begin{center}
       \includegraphics*[#6]{#2}
       \includegraphics*[#6]{#3}
       \includegraphics*[#6]{#4}
       \includegraphics*[#6]{#5}
       \caption{\label{#1}#7}
   \end{center}
 \end{figure}
}

\def\FIGss #1 #2 #3 #4 #5 #6 #7 [#8] #9\par{%
 \begin{figure}[!h]
   \begin{center}
       \includegraphics*[#8]{#2}
       \includegraphics*[#8]{#3}
       \includegraphics*[#8]{#4}
       \includegraphics*[#8]{#5}
       \includegraphics*[#8]{#6}
       \includegraphics*[#8]{#7}
       \caption{\label{#1}#9}
   \end{center}
 \end{figure}
}

\begin{document}

\title{Thermonuclear explosions of rapidly rotating white dwarfs --
I. Deflagrations} 

\author{J.M.M.~Pfannes\inst{1}, J.C.~Niemeyer\inst{1,2},
  W. Schmidt\inst{1,2} and C. Klingenberg\inst{3}}                      
 \authorrunning{J.M.M.~Pfannes, J.C.~Niemeyer, W.~Schmidt and C. Klingenberg}         
   \offprints{J.C.~Niemeyer}

\titlerunning{Deflagrations of rapidly rotating white dwarfs}
\authorrunning{J.M.M.~Pfannes et al.}

   \institute{Lehrstuhl f\"ur Astronomie,
Universit\"at W\"urzburg, Am Hubland, D-97074 W\"urzburg, Germany \\
             \email{pfannes@astro.physik.uni-wuerzburg.de}
             \and
             Institut f\"ur Astrophysik, Universit\"at G\"ottingen, Friedrich-Hund-Platz 1,
             D-37077 G\"ottingen, Germany\\
             \email{[niemeyer;schmidt]@astro.physik.uni-goettingen.de}
             \and Institut f\"ur Angewandte Mathematik,
Universit\"at W\"urzburg, Am Hubland, D-97074 W\"urzburg, Germany \\
             \email{klingenberg@mathematik.uni-wuerzburg.de}
             }
   \date{the date of receipt and acceptance should be inserted later}

   \abstract
       %
       %% Here the context to be.. 
       {Turbulent deflagrations of Chandrasekhar mass White Dwarfs are
       commonly used to model Type Ia Supernova explosions. In this
       context, rapid rotation of the progenitor star is plausible but
     has so far been neglected.} 
       %
       %% Here the aims to be.. 
       {The aim of this work is to
       explore the influence of rapid rotation on the deflagration scenario.} 
       %
       %% Here the methods to be.. 
       {We use three dimensional hydrodynamical simulations
       to model turbulent deflagrations ignited within a variety of rapidly
       rotating CO WDs obeying rotation laws suggested by accretion studies.} 
       %
       %% Here the results to be.. 
       {We find that rotation has a significant impact on the explosion. The
       flame develops a strong anisotropy with a preferred direction towards the stellar poles,
       leaving great amounts of unburnt matter along the equatorial
       plane.} 
       %
       %% Here the conclusions to be.. 
       {The large amount of unburnt matter is contrary to observed spectral
       features of SNe~Ia. Thus, rapid rotation of the progenitor star and the
       deflagration scenario are incompatible in order to explain SNe~Ia.}

 
\maketitle

%%%%%%%%%%%%%%%%%%%%%%%%%%%%%%%%%%%%%%%%%%%%%%%%%%%%%%%%%%%%%%%%%%%%%%%%%
%%%%%%%%%%%%%%%%%%%%%%%%%%%%%%%%%%%%%%%%%%%%%%%%%%%%%%%%%%%%%%%%%%%%%%%%%
%%%%%%%%%%%%%%%%%%%%%%%%%%%%%%%%%%%%%%%%%%%%%%%%%%%%%%%%%%%%%%%%%%%%%%%%%

\section{Introduction}
\label{intro}

Driven by the outstanding potential of Type Ia Supernovae (SNe Ia) to
reconstruct the expansion history of the late universe, a number
of observational campaigns have collected a wealth of empirical data
about the lightcurves, spectra, and host populations of these powerful
events \citep{essence,hstgold}.
This progress is in contrast to the sobering fact that despite many
decades of theoretical and computational efforts, we are still lacking
a solid understanding of the explosion mechanism, or the combination
of different mechanisms, that can explain the class of SNe Ia in its
entirety. While the majority of SN Ia explosions is believed to involve the
rapid thermonuclear combustion of Chandrasekhar mass CO White Dwarfs (WDs)
\citep{HillebrandtNiemeyer00,2008arXiv0804.2147R}, it is still debated exactly
how and 
where the star ignites
\citep{2004ApJ...607..921W,2006A&A...446..627S,2006A&A...448....1R},  
and whether the combustion front 
propagates subsonically as a deflagration or supersonically as a detonation
\citep{GCDmodel,snob,RoepkeNiemeyer07}. However, it is clear that
if the front is subsonic, its effective propagation rate is governed
only by the turbulence produced by large-scale instabilities
\citep{1997ApJ...475..740N}.

From a purely theoretical point of view, the turbulent deflagration model
is favoured compared with its competitors that invoke a spontaneous
transition to a detonation \citep{1999ApJ...523L..57N}, but large 3D
simulations 
indicate it may be incapable to explain the full class of SNe Ia
\citep{snob}.
Nevertheless, because of its conceptual
simplicity and 
lack of free parameters, we will use it as a testbed for studying the
influence of rapid rotation on the outcome of the explosion. Since the
duration of thermonuclear burning is determined by the dynamical
time scale of the WD, the star must rotate nearly critically in order
for rotation to have any significant impact. This appears plausible at
first sight, since most progenitor scenarios for SNe Ia involve the
accretion of almost a solar mass of material from a binary companion
before the thermonuclear runaway, allowing the WD to pick up a
substantial amount of angular momentum. Indeed, some evolutionary
calculations predict critical rotation of the WD at the time of the
explosion \citep{2005A&A...435..967Y}. On the other hand, the transport and
dissipation of angular momentum inside the star are poorly understood
and direct observations are absent. Hence, we treat the
amplitude and shape of the rotation law as essentially undetermined 
and attempt to investigate their influence in a parameterised
manner. As we will show, our general conclusion is robust with respect
to the details of the rotation law: nearly critical rotation of the
progenitor WD together with a pure deflagration model for the
combustion front are inconsistent with the observations of SNe Ia.

The first multidimensional simulations of rapidly rotating exploding
WDs were carried out by \citet{1992A&A...254..177S}. However, in
contrast to our work they
concentrated on pure detonations instead of deflagrations. We will
revisit the prompt detonation scenario in a follow-up paper. 

This paper is structured as follows. 
Section 2 summarises the theoretical background on rotation of SNe~Ia progenitor stars. The
motivation of the employed rotation laws is given.
Different methods for the numerical initiation of the burning process are presented in
section 3.  
Section 4 describes the computational grid, the flame
modelling, treatment of the gravitational potential, the employed WD models,
numerical stability, and the parameter study.
The results of the numerical study are listed in section 5. The impact of
rotation on the explosion is explained by considering buoyancy effects. Also,
the influence of model parameters and shear motion caused by rotation is
presented. An interpretation with respect to spectral features finalises the
section. 
Section 6 concludes the paper.

%%%%%%%%%%%%%%%%%%%%%%%%%%%%%%%%%%%%%%%%%%%%%%%%%%%%%%%%%%%%%%%%%%%%%%%%%
%%%%%%%%%%%%%%%%%%%%%%%%%%%%%%%%%%%%%%%%%%%%%%%%%%%%%%%%%%%%%%%%%%%%%%%%%
%%%%%%%%%%%%%%%%%%%%%%%%%%%%%%%%%%%%%%%%%%%%%%%%%%%%%%%%%%%%%%%%%%%%%%%%%

\section{Rotation of CO white dwarfs}
\label{wd_rot}

In the context of the single degenerate progenitor scenario for
SNe~Ia \citep{Liv00}, a white dwarf star consisting of carbon and oxygen (CO
WD) 
accretes matter from a non-degenerate companion 
via a Keplerian disc. Since the average WD mass is $\sim 0.6~M_{\sun}$,
the amount of 
matter that has to be accreted prior to C ignition is large
(this occurs soonest for non-rotating CO WDs close to their
Chandrasekhar mass, i.e. at $M_{WD} \sim 1.4~M_{\sun}$). 
In addition to the matter itself, angular momentum is
accreted. Therefore it is reasonable to assume that the WD spins up during
its accretion phase. 

\citet{2004A&A...419..623Y} (YL) analyse
the physics of accreting white dwarfs (AWDs) in detail and claim that in 
case of rapid rotation, the
limiting mass for C ignition or collapse can greatly exceed the
value of the Chandrasekhar-mass \citep{2005A&A...435..967Y}.
YL's stellar evolution code includes the effects of 
accretion-induced heating and energy transport, angular momentum transport by
various instabilities 
and the effect of rotation on the WD
structure by modifications in the momentum and energy conservation
equations. Electron and ion viscosities, relevant for the onset of
instabilities, are taken into account.
Transport of the accreted angular momentum is treated as a
diffusion process.

As their main conclusion regarding the angular velocity profile of the SN~Ia
progenitor, accreting WDs rotate differentially throughout their
evolution. The angular velocity reaches a maximum,  $\Omega_{\mathrm{peak}}$, 
within the star, increasing from the centre outwards but
decreasing toward the surface. This behaviour
is attributed to the fact that the slowly rotating inner part
contracts faster than the rapidly rotating surface layers as the total mass of
the accreting star increases. 
Its value is in the range of $1.5~\mathrm{rad/s} \lesssim
\Omega_{\mathrm{peak}} \lesssim 
6.5~\mathrm{rad/s}$ when the central density of the WD reaches 
the ignition density of $2.0 \times 10^{\,9}~\mathrm{g/cm^3}$.
The corresponding total masses 
are $\sim 1.5$ and $\sim 2.1~M_{\sun}$, respectively.

\begin{figure*}
\centering
\includegraphics[width=17cm]{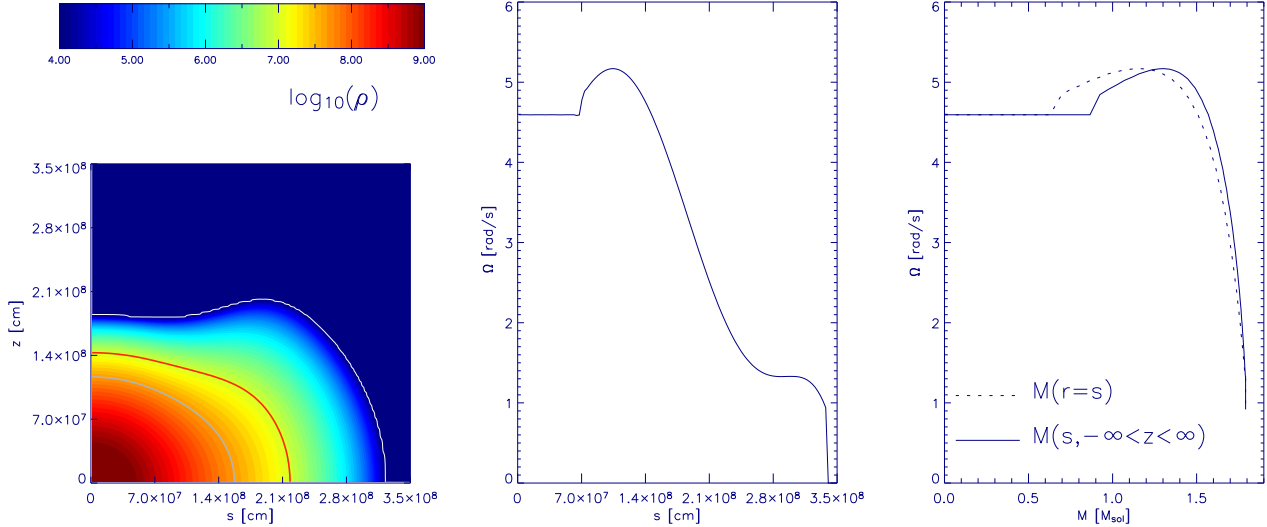}
\caption{Density contour lines (left panel) and the rotation law both in
  spatial (middle panel) and mass coordinates (right panel) for the AWD4
  model. Unlike the dotted line that 
  denotes the spherical mass coordinate, the solid line line indicates the
  cylindrical mass coordinate in order to describe cylindrical rotation.
  Note that the extent of the convective 
  core (i.e., the rigidly rotating inner part) is estimated following
  \citet{2004A&A...419..623Y}.}
\label{awd4}
\end{figure*}

As YL's results suggest, the rotation law(s) realised by nature
may depend on different factors such as efficiency of accreted angular momentum
transport, the accretion time scale, the time scale 
for loss of angular momentum, the total binary mass budget, or
magnetic fields. Therefore, in this work we study the impact of a
parameterised class of rotators
with different rotation laws and, consequently, different explosion
masses. Specifically, four types of AWD rotators are considered as
SN~Ia progenitors. They are 
-- listed in order of increasing stellar mass --
denoted by AWD1, AWD2, AWD4, 
and AWD3 and include the full spread of AWD masses
Physical quantities and numerical details are itemised in Table \ref{tab:ppa_coeffs}. 
The central density is $\rho_c = 2.0 \times 10^{\,9}$
g/cm$^3$ for all models. $M_{high}$ is the mass for all material if $\rho >
5.248 
\times 10^{\,7}~\mathrm{g/cm^3}$, and $M_{med}$ if $ 5.248
\times 10^{\,7}~\mathrm{g/cm^3} > \rho > 1.047 \times 10^7~\mathrm{g/cm^3}$. 
$\beta$ denotes the ratio of rotational energy and gravitational binding
energy, $J$ the angular momentum.
In the non-rotating case ($\rho_c = 2.0 \times 10^9~\mathrm{g/cm^3},$ $M =
1.4~M_{\sun}$) $M_{high}$ and $M_{med}$ are 1.268 $M_{\sun}$ (90.6 \%)
and 1.110 $M_{\sun}$ (7.9 \%), respectively. 
Furthermore, the non-rotating star possesses the following quantities:
$r = 2.1905 \times 10^{\,8}~\mathrm{cm}$,
$E_{int}=2.5228 \times 10^{\,51}~\mathrm{erg}$, 
$E_{grav}=-3.0275 \times 10^{\,51}~\mathrm{erg}$, $E_{rot}= 0.000
\times 10^{\,50}~\mathrm{erg}$, and
$E_{bind} \equiv E_{int} + E_{grav} + E_{rot} =-5.047 \times
10^{\,50}~\mathrm{erg}$.
Since rigid rotation is believed
to occur in the pre-supernova convective core due to very efficient exchange
of angular momentum, rigid rotation of the interior is assumed for the
AWD4 model (see Fig.~\ref{awd4}) while the rotators AWD1 to 3 rotate
differentially everywhere.   

Besides AWD rotation, different rotation laws have been tested for
pre-supernova WDs. If there are strong braking magnetic torques or
efficient viscous angular momentum transport, a critically rotating rigid rotator
might result. In contrast to differentially 
rotating WDs, the mass of this kind of object increases only 
slightly by $0.07~M_{\sun}$ for a central density $\rho_c =
2.0 \times 10^9~\mathrm{g/cm^3}$.
In addition, the frequently used $j-const.$ and $v-const.$ rotation laws,
denoting approximately constant specific angular momentum and rotational
velocity, respectively, are employed (cf. Table \ref{tab:j_v_const}). 
The parameter~$adr$ denotes the fraction of the equatorial radius up to
which rotation has approximately constant specific angular momentum and
rotational velocity, respectively \citep{1985A&A...146..260E}.
For both of these
rotation laws, angular velocity 
decreases steadily from the rotational axis outwards.
See section \ref{method} for details of the numerical
treatment of the employed rotating WDs.

Under certain restrictions, it is also possible to interpret our model in the context of the double-degenerate scenario for SNe Ia, i.e. the merger and immediate explosion of two WDs. Rapid rotation is particularly well motivated in this case. However, two conditions need to be satisfied for our model assumptions to apply: first, in order to reach hydrostatic equilibrium of the initial configuration, the explosion must only be ignited after a sufficiently long delay of several dynamical time scales. Second, it is unclear whether in this scenario, ignition will occur first near the surface and propagate inward (see \citet{HillebrandtNiemeyer00} for details and references). Note that while the former is simply a restriction on the applicability of our initial conditions, the latter is a serious problem for the double-degenerate scenario in itself. For the purpose of our study, we remain agnostic about the progenitor history as long as the conditions for viability of our initial model are met.

%%%%%%%%%%%%%%%%%%%%%%%%%%%%%%%%%%%%%%%%%%%%%%%%%%%%%%%%%%%%%%%%%%%%%%%%%
%%%%%%%%%%%%%%%%%%%%%%%%%%%%%%%%%%%%%%%%%%%%%%%%%%%%%%%%%%%%%%%%%%%%%%%%%
%%%%%%%%%%%%%%%%%%%%%%%%%%%%%%%%%%%%%%%%%%%%%%%%%%%%%%%%%%%%%%%%%%%%%%%%%

\section{Ignition conditions}
\label{ign_cond}

The explosion is assumed to be a turbulent deflagration
\citep{HillebrandtNiemeyer00}.
The structure of the flame surface at the time of ignition is known to have a strong impact on the
explosion outcome.
In order to investigate the impact of certain ignition situations, 
we used a variety of different methods to initiate the burning front in our simulations.  

A standard flame morphology that has been employed in several applications is the
$c3$~configuration \citep{2002A&A...386..936R}, representing a central
ignition with three bulges in azimuthal direction per octant. 
Besides the fact that the $c3$~ignition is reproducible 
and ensures similar conditions even if initiated in different initial
models,
a major advantage of the $c3$~ignition is the applicability for 
coarse numerical resolutions.

As a modification of the static multi-point ignition
\cite{2005A&A...432..969R}, the stochastic 
ignition model initiates flame seeds randomly at different instants in time
\citep{2006A&A...446..627S}.
The probability of an ignition depends on the background
temperature and the magnitude of the convective temperature fluctuations 
which are modelled by a mixing-length approach \citep{2004ApJ...607..921W,
2004ApJ...616.1102W}. Thus, the density of the emerging ignitions 
decreases with the distance from the centre
and becomes statistically symmetric in angular directions. Due to the
expansion of the 
white dwarf, the ignition probability decreases with time. The
only parameter 
of the model, the exponentiation parameter $C_{\mathrm{e}}$,  
controls the overall speed of the process and thus the total number of ignition
events.

An alternative scenario is the  {\it dipole jet flow}
\citep{2004ApJ...607..921W},  
where ignition sites are  
clustered along the rotation axis in one of the two 
hemispheres. The alignment of ignitions with the outward flow direction is
explained by the fact that the condition necessary for
ignition is met after stellar material has been transported through the hot
stellar centre by the dipole jet flow. However, the dipole flow --
also likely to occur in the non-rotating WD as a consequence of the convective pre-supernova
motion -- might be eliminated by a 
moderate amount of rotation \citep{2006ApJ...640..407K}.

%%%%%%%%%%%%%%%%%%%%%%%%%%%%%%%%%%%%%%%%%%%%%%%%%%%%%%%%%%%%%%%%%%%%%%%%%
%%%%%%%%%%%%%%%%%%%%%%%%%%%%%%%%%%%%%%%%%%%%%%%%%%%%%%%%%%%%%%%%%%%%%%%%%
%%%%%%%%%%%%%%%%%%%%%%%%%%%%%%%%%%%%%%%%%%%%%%%%%%%%%%%%%%%%%%%%%%%%%%%%%

\section{Numerical Method}
\label{method}

We use the hydrodynamics code {\sc Prometheus} \citep{Rei01} for the
simulation of explosions of rotating WDs in three dimensions. 
The computational grid is a moving grid composed of two
nested sub-grids \cite{2005A&A...432..969R}. The inner grid
captures the burned region and the flame surface, while the outer grid
expands together with the expanding WD. 
A crucial parameter for numerical simulations of SNe~Ia is flame propagation velocity, because
the evolution of the flame front and, consequently, the energy
release from thermonuclear burning is greatly enhanced by turbulence.
Following  \citet{2006A&A...450..265S} and
\citet{2006A&A...450..283S},
the speed of the flame front is computed by a
subgrid scale model that is based on a balance law for the
turbulent energy.

While previous simulations were performed with a gravitational
potential that is spherically symmetric, we employ a multipole solver
for the computation of the gravitational potential of rotating stars.
A parameter study concerning different expansion contributions in rotating
WDs showed that $\Phi_g(\vec{r})$ is approximated with sufficient
accuracy  if the axisymmetric multipole expansion is truncated after the quadrupole term.
The numerical implementation was adapted from the 
public {\sc FLASH} code \citep{Fry00}. The FLASH implementation of the multipole solver is in turn adapted from
the original implementation by \citet{MuellerStein1995}.

Rotating density stratifications in hydrostatic equilibrium had to be
generated as initial conditions for the hydrodynamic simulations
(see Fig.~\ref{awd4} for an example). We constructed initial models
using the method of \citet{1985A&A...146..260E}. This method, 
which is an iterative procedure of correcting the
density field based on local deviations from hydrostatic equilibrium,
guarantees the creation of 
stars in hydrostatic equlilibria for a broad range of rotation laws.  
The WD equation of state was approximated 
using the piecewise polytropic approximation method
\citep{1985A&A...152..325M}. Moreover, the rotators obey the
characteristic AWD rotation 
laws as shown by \citet{2004A&A...419..623Y}. This was achieved with
the implementation of a polynomial representation of the angular velocity 
(which was terminated after the $5^{\mathrm{th}}$~order;
cf. Table \ref{tab:ppa_coeffs} for the chosen coefficients):
\begin{align} \label{eq:rotlaw}
  \Omega(s) ~=~
  \begin{cases} ~\Omega_c &~\text{for} ~~~     0 \leq \bar s \leq \bar s_{\mathrm{rigid}} \\
    ~\Omega_c \;+\; c_1\, \bar s \;+\; c_2\, \bar s^2 \\ ~~~ \;+\; 
    c_3\, \bar s^3  \;+\; c_4\, \bar s^4 \;+\; c_5\, \bar s^5
    &~\text{for} ~~~    \bar s_{\mathrm{rigid}}< \bar s \leq \bar s_{\mathrm{max}}   \end{cases}
\end{align}
Here, $\bar s$ is the normalised distance from the rotation axis and 
$\bar s_{\mathrm{rigid}}$ the normalised distance from the rotation
axis out to the position for which the angular velocity has a constant
value of $\Omega_c$: 
\begin{align} \label{eq:s_norm}
  \bar s ~\equiv~ \frac{s}{r_{\mathrm{surf}}^{~~\mathrm{equator}}} 
~~~~~~\text{and}~~~~~~
  \bar s_{\mathrm{rigid}} ~\equiv~ \frac{s_{\mathrm{rigid}}}{r_{\mathrm{surf}}^{~~\mathrm{equator}}}
\end{align}
Since a strong magnetic field might
prevent the rapid rotation of accreting WD's
\citep{2004A&A...419..623Y}, and the 
possible shape of the rotation law may vary to a large degree for different
assumptions 
in the accretion model or simply from star to star, 
the initial models
employed for this study were 
chosen to cover a broad range of possible rotation
laws and, accordingly, masses. Besides the AWD rotation laws, we
also considered rigid
rotation as well as the $j-const.$ and $v-const.$ rotation laws, both stating
{\it decreasing} angular velocity outwards
(cf.\ \citet{1985A&A...146..260E} for details on both rotation laws).

Tables \ref{tab:ppa_coeffs} and
\ref{tab:j_v_const} summarise the parameters of our 
models. The masses of the AWD initial models range from $1.64~M_{\sun}$ to
$2.07~M_{\sun}$. Model AWD4 contains a rigidly rotating inner core (up to
the equatorial position $s = 6.76 \times 10^6$~cm) that is thought to arise
from the pre-supernova convective motion within the star. The rotation law for
this model is plotted in Fig.~\ref{awd4}.
Note that differential rotators can become heavier than the
critical rigid rotator since $\Omega$ is allowed to have a high value within
the star and to drop towards the equatorial surface. As a consequence,
overcritical rotation is avoided.

\begin{table}%[tp]
\begin{center}
\begin{tabular}{|@{~~}r@{~~}||*{4}{@{~\,}c@{~\,}|}}
\hline
model              &  AWD1  & AWD2 &  AWD4 & AWD3  \\
\hline\hline
$\Omega_c$~[rad/s] &  1.5999  & 4.6074 &  4.5934 & 3.9847  \\
\hline
$c_1$                      & 7.3470  & 4.5    &-16.265 &-18.823  \\
\hline
$c_2$                      & 62.006  & -15    & 162.39 & 196.62  \\ 
\hline
$c_3$                      & -275.25 & 9      &-480.84 &-563.38  \\ 
\hline
$c_4$                      & 342.62  & 0      & 531.85 & 600.25  \\ 
\hline
$c_5$                      & -137.09 & 0      &-2.0081 &-218.51  \\ 
\hline
$\bar s_{\mathrm{rigid}}$  & 0       & 0      &  0.2   & 0       \\
\hline
$r_{equator} / r_{pol} $ & 1.629 & 1.710 & 1.796 & 2.183   \\
\hline\hline
$\Omega_{peak}$~[rad/s]& 4.4126  & 4.5158  & 5.1986 & 5.2236 \\
\hline
$M ~[M_{\sun}]$      & 1.6374  & 1.7428 & 1.7911 & 2.0150  \\
\hline
$M_{high}~[M_{\sun}]$&  1.393  & 1.481  & 1.490 & 1.469  \\
\hline
$M_{high}/M ~[\%]$ &  85.1    & 85.0   & 83.2  & 72.9    \\
\hline
$M_{med}~[M_{\sun}]$ &   0.180  & 0.200  & 0.227   & 0.401 \\
\hline
$M_{med}/M ~[\%] $ &  11.0    & 11.5   & 12.7  & 19.9  \\
\hline
$r_{pol} ~[10^{\,8} \mathrm{cm}]$ & 2.0180 & 1.8985 & 1.8833  & 1.8836 \\
\hline
$ r_{equator} ~[10^{\,8} \mathrm{cm}]$  & 3.2871 & 3.2471  & 3.3822 & 4.1116 \\
\hline
$E_{int}~[10^{\,51}\mathrm{\,erg}]$  & 2.7395 & 2.8895  & 2.9148 & 2.9604 \\
\hline
$E_{grav}~[10^{\,51}\mathrm{\,erg}]$   & -3.6701 & -4.0330 & -4.1419 & -4.5283  \\
\hline
$E_{rot}~$[10$^{\,50}$\,erg]      & 1.748 & 2.625 & 2.991 & 4.465   \\ 
\hline
$E_{bind}~$[10$^{\,50}$\,erg]  & -7.558 & -8.810 &  -9.280  & -11.214 \\ 
\hline
$\beta~[\%]$                      & 4.7639 & 6.5092  & 7.2206 & 9.8608 \\
\hline
$ J~[10^{\,50}\,\mathrm{g \;cm}^2 / \mathrm{s}]$  & 0.91512 & 1.1968 & 1.3602 & 2.1951 \\
\hline
\end{tabular}
\end{center}
\caption{Physical quantities and numerical coefficients of the constructed AWD
  initial models.}
\label{tab:ppa_coeffs}
\end{table}

\begin{table}%[tp]
\begin{center}
\begin{tabular}{|r||*{3}{c|}}
\hline
model              & rigid  &  $j_{const}$  & $v_{const}$ \\
\hline\hline
$\Omega_c$~[rad/s] & 0.37242  & 6.4076  & 10.120   \\
\hline
$r_{equator} / r_{pol} $ & 1.477 & 2.183 & 2.183 \\
\hline
$adr$              &  -  &  0.5  & 0.2 \\
\hline\hline
$M ~[M_{\sun}]$ & 1.4661    & 1.8025  & 1.7067   \\
\hline
$M_{high}~[M_{\sun}]$ &  1.302          & 1.534 & 1.493    \\
\hline
$M_{high}/M ~[\%]$ &  88.8          & 85.1 & 87.5 \\
\hline
$M_{med}~[M_{\sun}]$ &  0.126         & 0.191 & 0.157     \\
\hline
$M_{med}/M ~[\%] $ & 8.6        & 10.6 & 9.2 \\
\hline
$r_{pol} ~[10^{\,8} \mathrm{cm}]$ &  2.1721         & 1.8087 & 1.8134 \\
\hline
$ r_{equator} ~[10^{\,8} \mathrm{cm}]$ &  3.2092        & 3.9481 & 3.95858 \\
\hline
$E_{int}~[10^{\,51}\mathrm{\,erg}]$  &  2.5878          & 3.0072 & 2.9678 \\
\hline
$E_{grav}~[10^{\,51}\mathrm{\,erg}]$ & -3.2074           & -4.2779 & -4.0604 \\
\hline
$E_{rot}~$[10$^{\,50}$\,erg]     &  0.442            & 3.163 & 2.393 \\ 
\hline
$E_{bind}~$[10$^{\,50}$\,erg]    &  -5.754             & -9.544 & -8.533 \\ 
\hline
$\beta~[\%]$                    &  1.3769             & 7.3941 & 5.8933 \\
\hline
$ J~[10^{\,50}\,\mathrm{g \;cm}^2 / \mathrm{s}]$ & 0.37747    & 1.3785 & 1.0535 \\
\hline
\end{tabular}
\end{center}
\caption{The same parameters as in Table \ref{tab:ppa_coeffs} for rigid
  rotation and the
  ``exotic'' $j_{const}$ and $v_{const}$ rotators.}
  \label{tab:j_v_const}
\end{table}

The rotation law was not fitted for an exact overlap with those
proposed by YL. In contrast, our parametrisation was chosen to emulate
the central angular velocity, $\Omega_c$,
the maximum of $\Omega$, and the total mass. Given the
uncertainties in the predictions for rotation laws, this choice is
sufficiently realistic for our purposes.

The mass accretion from a Keplerian disc implies a
critically rotating surface of the AWD rotators which cannot easily be
incorporated because the initial model algorithm aborts when mass
shedding for a model is detected. Therefore it is assumed that  
the surfaces of most constructed models in this study do not 
rotate critically. This is not problematic since the outer layers of a
star do not affect the nuclear burning in the interior.  

We tested whether the initial models remain hydrodynamically 
stable on the grid structure used for the SNe~Ia simulations.
Hydrodynamical simulations were followed to
$t=1.5~\mathrm{s}$, the time when burning would have stopped if the
WD had been ignited.
A comparison of density contours at different instants served as a
simple check for hydrostatic stability.
Up to $t=1.0~\mathrm{s}$, the contour lines retained their shape
except at the surface. At later times, the WD's core density also
decreases somewhat 
because of the limited feasibility to perform circular motion on a rectangular
cartesian grid. 
Two reasons are responsible for amplified 
deviations in the outer layers. First, the resolution of the grid is lower
outside the core region and second, the pressure gradient is neglected below the threshold density
of $\rho = 10^{-3}~\mathrm{g/cm^{\,3}}$.
A more quantitative test was done by looking at the radial velocities emerging
during the test of hydrostatic stability.
Radial velocities in the inner region are at all instants
found to be 
$\lesssim 4 \times 10^{\,7}~\mathrm{cm/s}$ and
therefore one order of magnitude lower than the characteristic
velocities after ignition.
In conclusion, stability of all discretised models on the grid can 
be assumed. 

\begin{figure*}[t]%[tp]
  \mbox{
    \subfigure[~$\vec{g}_{\rm{grav}}$] 
    {\includegraphics[width=8.5cm]{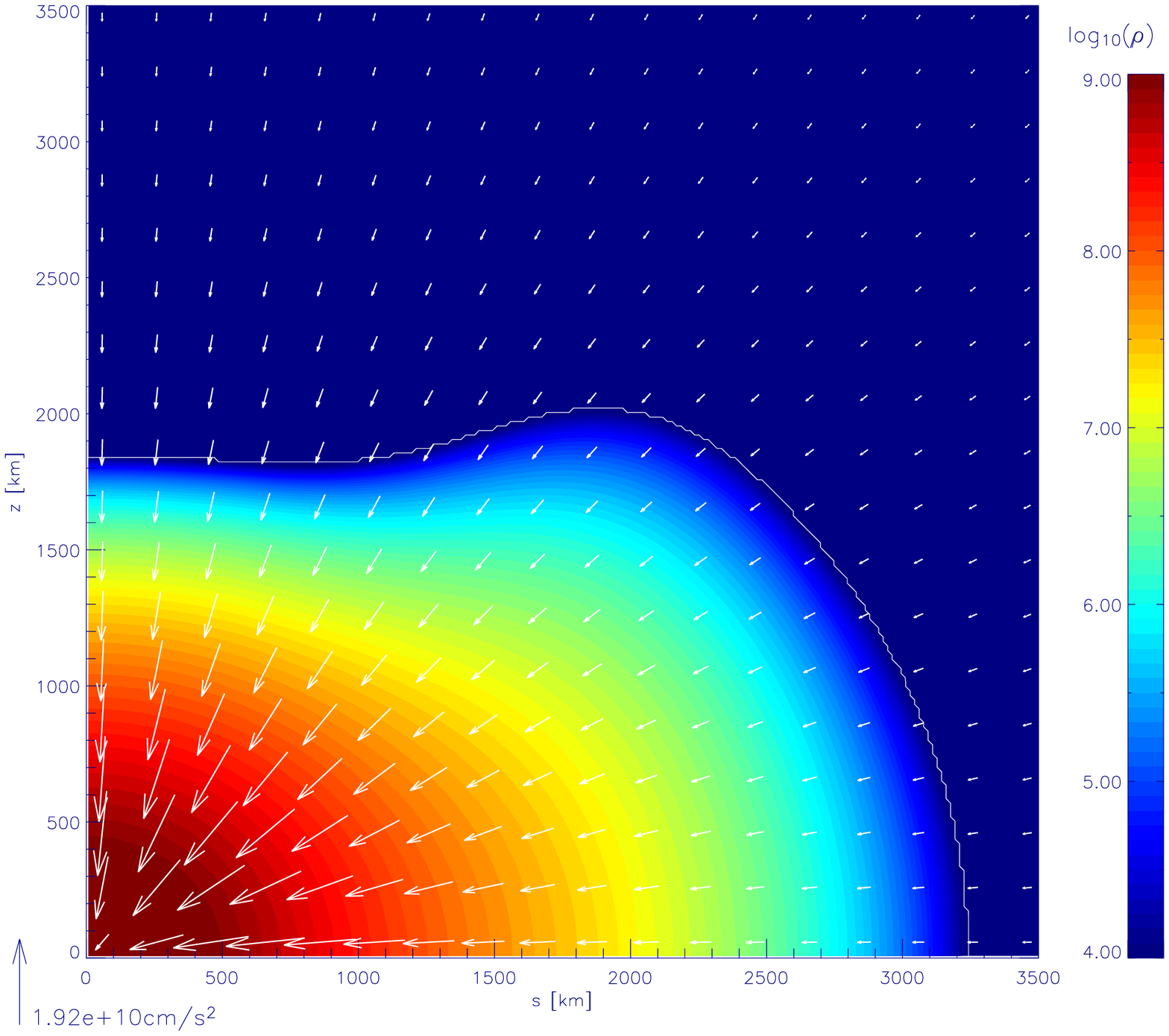}
      \label{fig:14AWD4_denstyggrav}}
    \subfigure[~$\vec{g}_{\rm{eff}}$] 
    {\includegraphics[width=8.5cm]{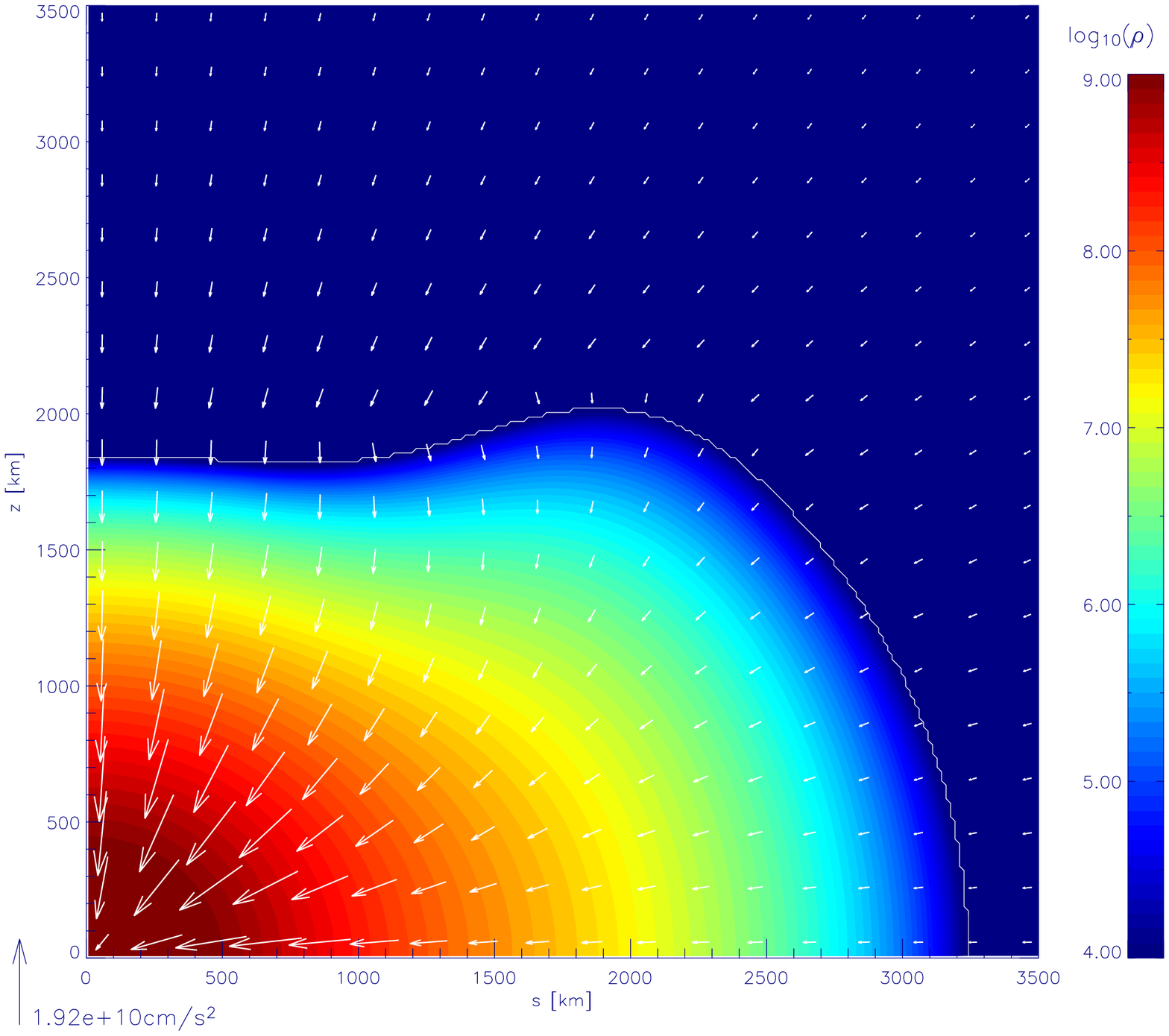}
      \label{fig:14AWD4_denstygeff}}}
    \caption{Density contour lines together with the gravitational acceleration
      $\vec{g}_{\rm{grav}}$ (a)
      and the effective gravitational acceleration
      $\vec{g}_{\rm{eff}}$ (b) that
      is indicated by arrows, respectively, for the  
      AWD4 rotator. Note that $\vec{g}_{\rm{eff}}$ is orthogonal to the
      coinciding isobaric and isopycnic surfaces (outside the star,
      the centrifugal acceleration is zero).}
\end{figure*}

\begin{figure}[t]%[p]
\resizebox{\hsize}{!}{\includegraphics{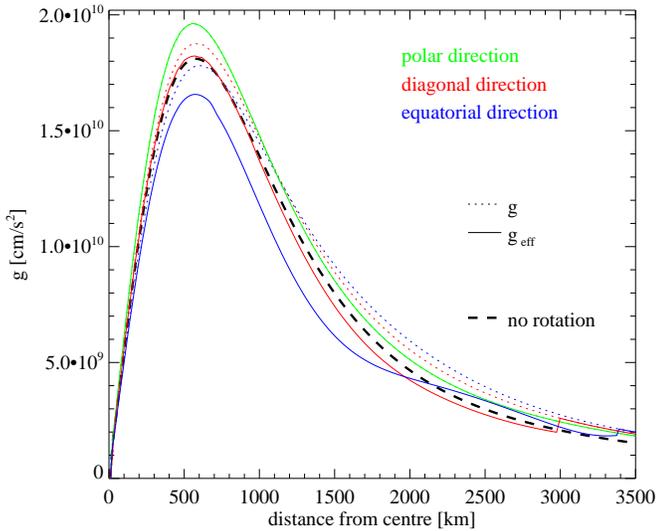}}
\caption{Gravitational acceleration $\vec{g}_{\rm{grav}}$ (dotted line) and
  the effective gravitational 
  acceleration $\vec{g}_{\rm{eff}}$ (solid line) along the rotation axis, the
  stellar diagonal, 
  and the equatorial plane for the AWD4 rotator. The edges of the red
  line (at $\sim 3000~\rm{km}$) and the blue line (at $\sim
  3400~\rm{km}$) appear due to the absence of centrifugal
  acceleration 
  outside the star. The dashed thick black line represents the spherically
  symmetric gravitational acceleration for the non-rotating star of
  equal central density.}
\label{fig:14AWD4_ggrav_geff_1D}
\end{figure}

In order to cover a broad range of possible
explosion scenarios, we varied the following parameters:
\begin{itemize}
\item the initial model: norot, rigid,
  AWD1, AWD2, AWD4, AWD3, $j_{const}$, and $v_{const}$
\item the ignition scenario:
  \begin{itemize}
    \item static ignition ($c3$)
    \item stochastic ignition; here, the ignition speed was varied by
    changing the parameter $C_e$ (cf. section \ref{ign_cond})
  \end{itemize}
\end{itemize} 
All simulations for this work were carried out in three spatial
dimensions. At least one hemisphere or even a full
star (necessary for the simulation of the dipole jet flow ignition) were
considered. Since our objective was a parameter study, we 
chose a relatively coarse initial resolution of
$7.5 \times 10^{\,5}~\mathrm{cm}$ on a $256^{2}\times128$ grid
for one hemisphere.
In order to get an estimate of the accuracy of all simulations, the 
``rigid $C_e = 5 \times 10^{\,4}$\,'' 
explosion model was redone with doubled resolution. 
It was in remarkable agreement with the run on a normal grid
(the explosion energetics differ by less than 4\%, the produced species differ
by less than 2\%). 
This result indicates that the resolution
employed for most of the simulations is sufficient
since the explosion energetics and species composition are almost identical.

%%%%%%%%%%%%%%%%%%%%%%%%%%%%%%%%%%%%%%%%%%%%%%%%%%%%%%%%%%%%%%%%%%%%%%%%%
%%%%%%%%%%%%%%%%%%%%%%%%%%%%%%%%%%%%%%%%%%%%%%%%%%%%%%%%%%%%%%%%%%%%%%%%%
%%%%%%%%%%%%%%%%%%%%%%%%%%%%%%%%%%%%%%%%%%%%%%%%%%%%%%%%%%%%%%%%%%%%%%%%%

\section{Results}
\label{results}

In the following, we will first analyse deflagration for the AWD4 rotator, which is ignited by
stochastic ignition. The strong impact of rotation on the flame morphology 
can be regarded as typical for the deflagration of rapidly rotating WDs. 
Variations of the stochastic ignition process and the influence of the rotation laws are investigated next.
We will also comment on the role of shear motion and the expected spectral features.  

\subsection{Anisotropic burning induced by rotation}
\label{impact_buoyancy}

The existence of a preferred direction for the flame propagation in rotating white dwarfs can be
attributed to two effects. First, the ashes of SNe~Ia are subject to buoyant motion, which is stronger in polar direction. Second, for convective motion in equatorial direction, the burning products 
have to gain angular momentum, whereas the opposite is true for inflowing fuel.
This effectively gives rise to an angular momentum barrier,
which suppresses the turbulent spreading of the deflagration front in the direction
orthogonal to the rotation axis. 

Fig. \ref{fig:14AWD4_denstyggrav} shows 
the gravitational acceleration $\vec{g}_{\rm{grav}}$ inside the
AWD4 rotator. As a result of the larger density gradient
along the rotation axis, $\vec{g}_{\rm{grav}}$ is stronger
(indicated by longer arrays) in this direction compared to the
equatorial plane. Buoyancy is caused by the effective
gravitational acceleration $\vec{g}_{\rm{eff}}$, i.e., the
gravitational acceleration including the
centrifugal acceleration as plotted in
Fig. \ref{fig:14AWD4_denstygeff}. $\vec{g}_{\rm{eff}}$ is orthogonal to 
the coinciding isobaric and isopycnic surfaces
and exactly compensates the 
pressure gradient in hydrostatic equilibrium. The difference in
$\vec{g}_{\rm{eff}}$ along the polar
and equatorial direction is even larger than the difference in
$\vec{g}_{\rm{grav}}$, which can be seen in
Fig. \ref{fig:14AWD4_ggrav_geff_1D}. 

\begin{figure*}[t]
\centering
\includegraphics[width=17cm]{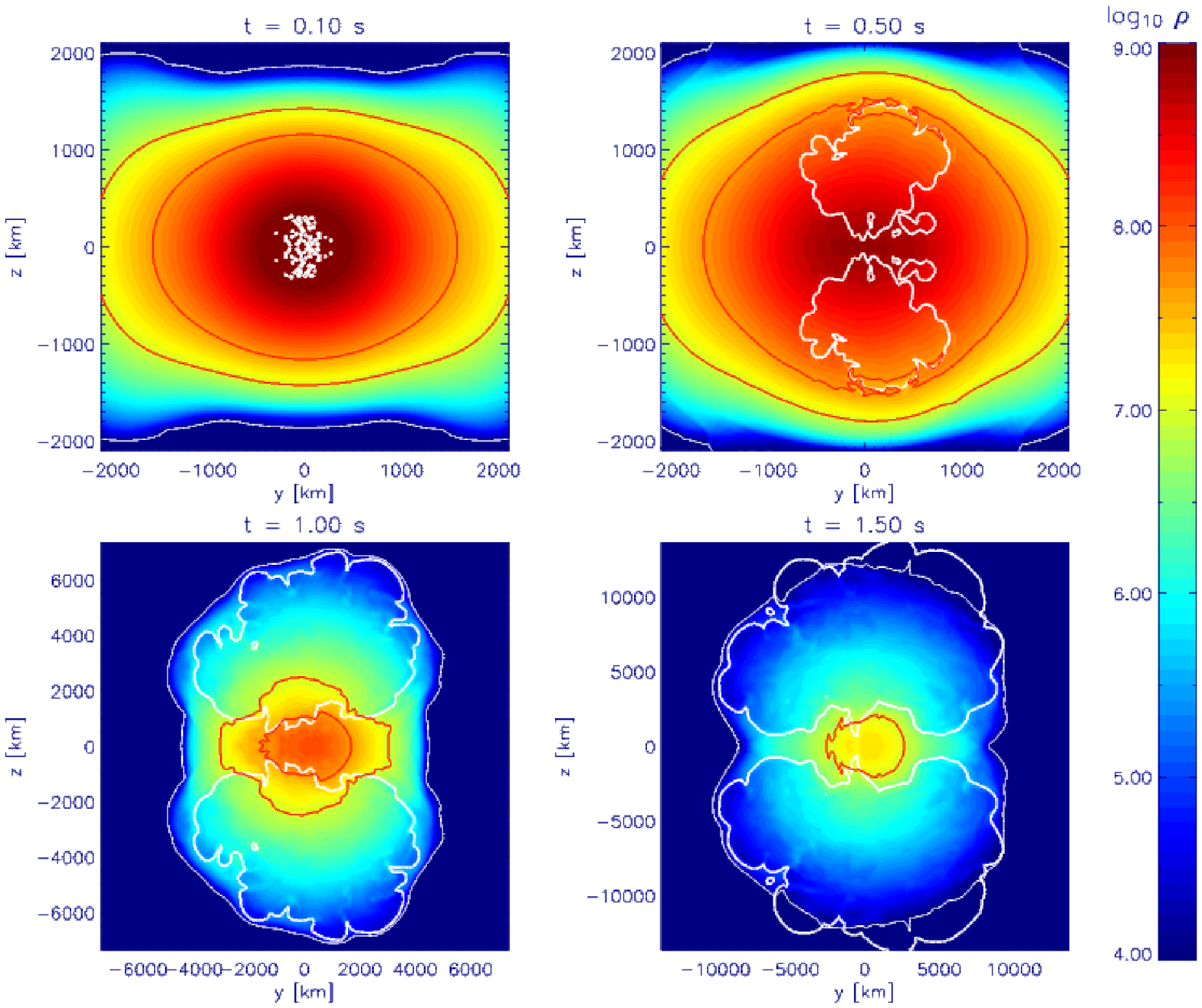}
\caption{Density contour lines for the pure deflagration of the
AWD4 rotator at different instants
for stochastic ignition ($C_e = 5 \times 10^{\,4}$). Cross sections
along the rotation axis of these 
simulations exhibiting equatorial symmetry are shown.}
\label{fig:14AWD4_stoch}
\end{figure*}

For the velocity $v_b$ of a buoyant bubble, the following expression
holds \citep{1950RSPSA.200..375D,1988PhFl...31.2077G, 1984PhyD...12...45R}:  
\begin{equation} \label{eq:vrise}
  v_b ~=~ C_{\,1}\sqrt{\frac{1}{2}\mathrm{At}\,g_{\rm{eff}}D},
\end{equation}
where the dimensionless constant $C_1$ is about 0.5, $D$ is the bubble
diameter, and the Atwood number $\mathrm{At}=(\rho_{\mathrm{f}}-\rho_{\mathrm{b}})/(\rho_{\mathrm{f}}+\rho_{\mathrm{b}})$ specifies the density contrast between the
rising burning products of density $\rho_{\mathrm{b}}$ in fuel of
density $\rho_{\mathrm{f}}>\rho_{\mathrm{b}}$.
The dependence of $v_b$ on the position inside the star is mainly given
by the variation of $\vec{g}_{\rm{eff}}$. The magnitude of
$\vec{g}_{\rm{eff}}$ differs by a factor up to
$1.3$ in polar and equatorial directions. 
(see Fig. \ref{fig:14AWD4_ggrav_geff_1D}). 
From equation~(\ref{eq:vrise}) follows that the rising velocity of ashes
in polar direction can exceed the velocity in equatorial direction by 15~\%. 
The velocity of the fastest bubbles is of the order $\sim 10^{\,8}~\rm{cm/s}$ (with $\rm{At}
= 0.5$, $|\vec{g}_{\rm{eff}}| = 2 \times
10^{\,10}~\rm{cm/s^{\,2}}$, and a bubble diameter of
$10^{\,8}~\rm{cm}$), a value that is approximately 10~\% of the
speed of sound. Accordingly, smaller bubbles rise more slowly. A lower limit
for the rising speed is of the order of $10^{\,6}~\rm{cm/s}$ (with $\rm{At}
= 0.1$, a lower gravitational acceleration and the bubble dimension of
$10^{\,5}~\rm{cm}$).

In the following, we consider the explosion of the AWD4 rotator in the stochastic ignition scenario
with $C_e = 5 \times 10^{\,4}$. 
Fig. \ref{fig:14AWD4_stoch} shows the flame surface (represented by thick 
white contours) embedded in density profiles at four instants.
The simulation was carried out for one hemisphere only. Thus the contour plots --- cross sections
in the $yz$-plane along the rotation axis --- exhibit equatorial symmetry.
During the first $0.1$ seconds, 363 ignition spots are set 
isotropically in the interior of the star. Even at this early period, an  
agglomeration of the flame around the rotation axis (which points
along the $x$ axis) is visible. Subsequently, the anisotropy of the flame
grows because of the effects discussed above. In addition, the turbulent
flame propagation speed is amplified in the polar direction due to the
stronger production of turbulence. The thin contour lines (red in the online version) in Fig. \ref{fig:14AWD4_stoch} enclose densities greater than $5.248 \times 10^{\,7}~{\rm g/cm^3}$
(threshold for producing iron group elements) and $1.047 \times 10^{\,7}~{\rm g/cm^3}$ (threshold for producing intermediate mass elements).

In conclusion, both the buoyancy that is enhanced at the poles in
rotating WDs and the blocked 
mixing orthogonal to the rotation axis
lead to a strongly anisotropic flame 
that leaves behind unburnt material at the stellar centre. 

\subsection{Ignition speed in the stochastic ignition scenario}

As outlined in section \ref{ign_cond}, the stochastic ignition process
contains one free parameter: the exponentiation parameter $C_e$ that
controls the overall speed of the creation of
ignitions. Just as in the original work by \citet{2006A&A...446..627S},
this parameter was varied over a broad range in order to study the
influence of $C_e$. For this reason, the AWD3
rotator was ignited for $C_e$  
between $5\times10^{\,1}$ and $5\times10^{\,7}$
(see Table \ref{tab:sntd_stochignt_20AWD3}) in the isotropic mode,
i.e., for ignitions that are placed without preference. 
$\Omega$ denotes the total spherical angle covered by the simulation domain. 

\begin{table}%[t]%[tp]
\begin{center}
\begin{tabular}{|@{\,}r@{\,}||*{6}{@{\,}c@{\,}|}}
\hline
$C_e$                       & $5\times10^{\,1}$ & $5\times10^{\,3}$ &$5\times10^{\,4}$ & $5\times10^{\,5}$ & $5\times10^{\,7}$ & $10^{\,4}$ \\
\hline
mode                        & iso    & iso    & iso    & iso    & iso    & dipole \\ 
\hline
$\Omega~[\pi]$              &   2    &   2    &    2   &   2    &   2    &   4    \\   
\hline
$I_{2\pi}$                  & 13     & 152    & 391    & 809    & 19980  & 247    \\
\hline\hline
\multicolumn{7}{|l|} {t~=~5\,s} \\
\hline
$E_{kin}~$[10$^{\,50}$\,erg]  & 2.87  & 3.61    & 5.30   & 6.90   & 6.11   & 4.43  \\
\hline
$E_{tot}~$[10$^{\,50}$\,erg]  & 0.21  & 1.33    & 3.54   & 5.47   & 4.62   & 2.55   \\
\hline
$E_{nuc}~$[10$^{\,51}$\,erg]  & 1.050 & 1.182   & 1.420  & 1.619  & 1.535  & 1.310  \\
\hline
IGEs~$[M{\sun}]$         & 0.40  & 0.51    & 0.69   & 0.79   & 0.74   & 0.62   \\
  $[\%\, M_{tot}]$      & 20    & 25      & 34     & 39     & 37     & 31     \\
\hline
IMEs~$[M{\sun}]$         & 0.51  & 0.46    & 0.41   & 0.46   & 0.45   & 0.40   \\
  $[\%\, M_{tot}]$      & 25    & 23      & 20     & 23     & 23     & 20     \\
\hline
C+O~$[M{\sun}]$         & 1.10  & 1.04    & 0.91   & 0.76   & 0.82   & 0.99   \\
  $[\%\, M_{tot}]$      & 55    & 52      & 46     & 38     & 40     & 49     \\
\hline
\end{tabular}
\end{center}
\caption{Energetics and compositions for the deflagration models activated by stochastic ignition with different values for $C_e$ in
  the AWD3 rotator ($M_{tot} = 2.02~M_{\sun}$).}
  \label{tab:sntd_stochignt_20AWD3}
\end{table}

As summarised in Table \ref{tab:sntd_stochignt_20AWD3}, increasing $C_e$ results in a higher
amount of iron group elements (IGEs), while the production of intermediate mass elements (IMEs)
does not change significantly. However, similar to the results obtained by
\citet{2006A&A...446..627S}, the energy output and the IGE production
do not increase further for $C_e > 5\times10^{\,5}$. This effect can be explained by the
saturation or even decrease of the flame surface during the ignition phase  
for an extremely higher number of ignition events
(see \citet{2006A&A...448....1R} for 
a discussion of the same trend for the static multi-spot
ignition).
However, employing the value $C_e = 5\times10^{\,5}$ amounts to
burning the entire core almost instantaneously because within the
first tenth of a second almost $10^3$ ignitions are initiated within a 
central region of the typical radius of $\sim
350~\mathrm{km}$. Since high values of $C_e$ not only
seem unphysical but also cannot serve as a remedy for the problems
arising for the stochastic ignition in combination with rotation
(cf. section \ref{rotlaw_stochignt}), the 
values of $C_e = 5\times10^{\,3}$ and $C_e = 5\times10^{\,4}$ were used
in this work. 

The dipole jet flow scenario
was also tested by means of the 
AWD3 rotator. However, as 
the impact of buoyancy
already suggests (cf. section \ref{impact_buoyancy}), this ignition
realisation does not ameliorate the problem of remaining fuel close to the
centre. If ignitions are set mainly in one hemisphere (without loss of
generality, the southern one), the northern
hemisphere is also burnt at a later time, but C and O are left over in the
core.
In the non-rotating case,
a strong off-centre ignition also leads to a delayed burning of the distant
stellar part, but less centrally located fuel remains \citep{2007ApJ...660.1344R}.

We also found that the detailed location of randomly chosen ignition points
has a significant impact on the outcome of the simulation.
The ``rigid $C_e = 5 \times 10^{\,3}$\,'' explosion model 
was simulated twice and is summarised in the two colums $rigid$ and $rigid~^*$ of
Table~\ref{tab:sntd_rotlaw_stochignt}.
The
only difference was a different seed for the random number generator
which sets the ignition position within a shell whose number is determined by
the 
Poisson process. Due to the randomness of the ignition process, the
total 
number of ignitions is not entirely but approximately the same: 167 vs. 163
ignitions 
were set during the simulations. However, since 
bubbles inserted close to the rotation
axis grow much more rapidly and tend to dominate the explosion,
the generated nuclear energy differs by $\sim 20~\%$
%
% NEW_Pentecost_6
%
(the more fuel remains unburnt, the lower is the value of produced nuclear energy). 
This demonstrates the high
sensitivity of the rotating WD deflagration scenario on initial conditions.
The sensitivity to the initial distribution of the seed flame was also
observed for non-rotating WDs but in that context it caused only small deviations
\citep{2006A&A...446..627S}. Any stellar rotation enhances this finding as a
result of the oblate shape of the rotators. 

\begin{figure*}[t]
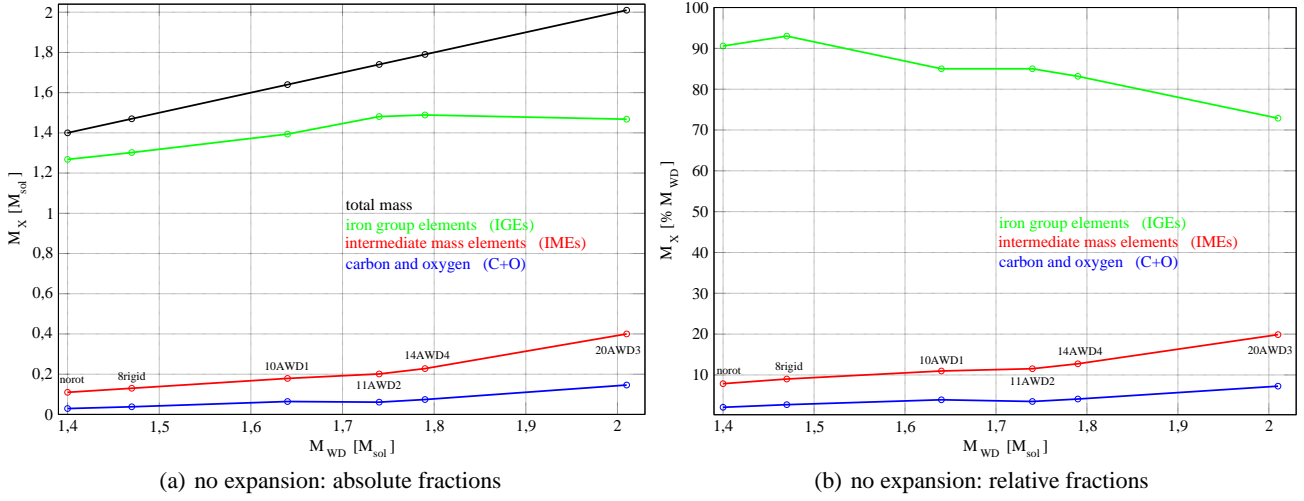
%[tp]
    \mbox{
    \subfigure[no expansion: absolute fractions] 
    {\includegraphics[width=8.5cm,clip]{2032fig5a.eps}
      \label{fig:composition_noburn_abs}}
    \subfigure[no expansion: relative fractions]
    {\includegraphics[width=8.5cm,clip]{2032fig5b.eps}
      \label{fig:composition_noburn_rel}}}
    \caption{Estimate of the burning products if the stars were not
      expanding during the burning process. The figure shows the fractions of the total mass in the high density
      regime ($\rho > 5.25 \times 10^{\,7} \mathrm{g/cm^3}$, ``IGEs''), the
      medium density regime ($ 5.25 \times 10^{\,7} \mathrm{g/cm^3} >
      \rho > 1.05 \times 10^7~\mathrm{g/cm^3}$, ``IMEs''), and the low
      density regime ($\rho \leq 1.05 \times 10^7~\mathrm{g/cm^3}$ ,
      ``C+O''). }
\end{figure*}

\begin{figure*}[t]
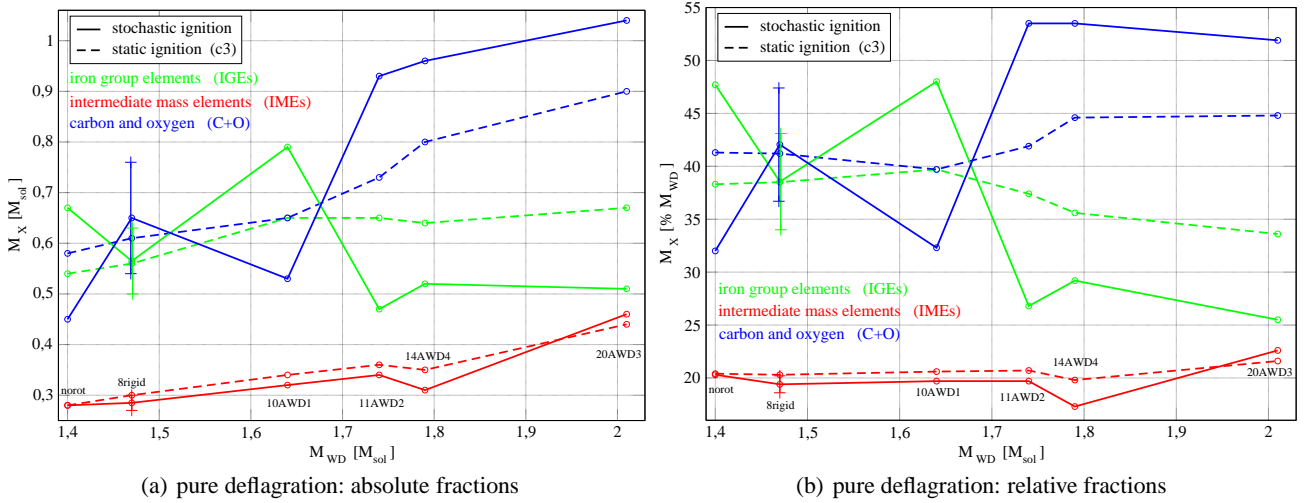

    \mbox{
    \subfigure[pure deflagration: absolute fractions]
    {\includegraphics[width=8.5cm,clip]{2032fig6a.eps}
      \label{fig:composition_burn_abs}}
    \subfigure[pure deflagration: relative fractions]
    {\includegraphics[width=8.5cm,clip]{2032fig6b.eps}
      \label{fig:composition_burn_rel}}}
    \caption{Burning products after $t = 5~\mathrm{s}$ as the outcome
      of simulations that are ignited by stochastic
      ignition (solid lines) and $c3$~ignition (dashed
      lines). Note the influence of the stochastic burning on the
      explosion outcome as can be seen for the rigid rotator: there the
      burning products are taken as the mean of two
      identical simulations that differ only in the random numbers for
      stochastic ignition.}
\end{figure*}

\subsection{Variation of the rotation law} \label{rotlaw_vari}

Differences in the rotation law give rise to substantial variations of the total
mass of the thermonuclear supernova progenitor. In particular, 
rapidly rotating WDs can greatly exceed the
canonical Chandrasekhar mass: the progenitor stars under consideration
possess masses in the range of $1.40~M_{\sun}$ (no rotation) and
$2.02~M_{\sun}$ (model AWD3).
An important 
conclusion can already be drawn by looking at the mass densities which
the different WD models exhibit. The total mass increases by $\sim 44~\%$
from the non-rotating star to the AWD3 rotator. However, the mass fraction with
densities greater than $5.25 \times 10^{\,7}~\mathrm{g/cm^3}$, for which
IGEs are produced, grows by only $\sim 16~\%$. 
The reason is that the fraction of
dense material within the star decreases from $\sim 91~\%$ in the non-rotating
star to only $\sim 73~\%$ in the AWD3 rotator which --- although heavier
by $\sim 0.6~M_{\sun}$ --- contains only $\sim 0.2~M_{\sun}$ more
IGEs. This is illustrated in Figs. \ref{fig:composition_noburn_abs} and
\ref{fig:composition_noburn_rel}. Therefore it is doubtful whether
the pure deflagration of rapidly rotating super-Chandrasekhar-mass
models yields significantly more $^{56}\mathrm{Ni}$ compared to the
non-rotating case. We consider the different rotation laws for
the $c3$ and stochastic ignition scenarios in turn.

\begin{figure}[th]
  \begin{minipage}{0.5\textwidth}
    \centering
   \subfigure[rotation law: $c3$~ignition] 
	      {\includegraphics[width=\linewidth,height=0.65\linewidth]{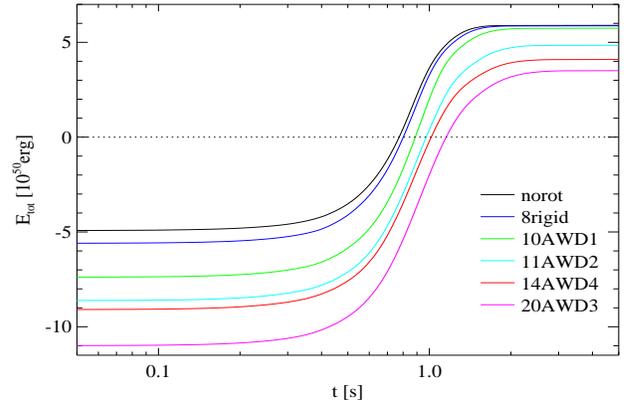}
		\label{panel_etot_c3}}
    \subfigure[rotation law: stochastic ignition]
	      {\includegraphics[width=\linewidth,height=0.65\linewidth]{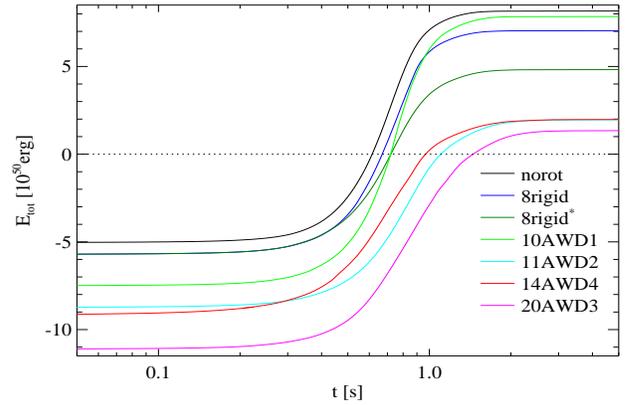}
		\label{panel_etot_stoch}}
    \caption{Temporal evolution of the total energy for the deflagration study that is
      initiated by $c3$~ignition in different rotators
      (a), and by stochastic ignition in different rotators ($C_e = 5
      \times 10^{\,4}$ for the AWD4 rotator, else $C_e = 5 \times 10^{\,3}$)
      (b)}
    \label{fig:sntd_etot}
  \end{minipage}\hfill
 \end{figure}

\begin{table}[t]%[tp]
\begin{center}
\begin{tabular}{|@{\,}r@{\,}||*{6}{@{\,}c@{\,}|}}
\hline
                           & norot & rigid & AWD1 & AWD2 & AWD4 & AWD3 \\
\hline\hline
$M_{tot}~[M{\sun}]$          & 1.40  & 1.47   & 1.64   & 1.74   & 1.79   & 2.02   \\
\hline\hline
$\Omega~[\pi]$               &   2   &   2     &    2  &   2    &   2    &   2    \\   
\hline\hline
\multicolumn{7}{|l|} {t~=~0\,s} \\  
\hline
$E_{grav}~[10^{\,51}\mathrm{\,erg}]$ &-3.027 & -3.204 & -3.665 & -4.027 & -4.135 & -4.518  \\
\hline
$E_{rot}~$[10$^{\,50}$\,erg]  & 0.00  & 0.44   & 1.75   & 2.62   & 2.99   & 4.46   \\
\hline
$\beta~[\%]$            & 0.00   & 1.38   & 4.78   & 6.50   & 7.22   & 9.87   \\
\hline\hline
\multicolumn{7}{|l|} {t~=~5\,s} \\
\hline
$E_{kin}~$[10$^{\,50}$\,erg]  & 6.49  & 6.53 & 6.63   & 5.99   & 5.42   & 5.28   \\
\hline
$E_{tot}~$[10$^{\,50}$\,erg]  & 5.89  & 5.87   & 5.74   & 4.84   & 4.09   & 3.50   \\
\hline
$E_{nuc}~$[10$^{\,51}$\,erg]  & 1.076 & 1.129  & 1.291  & 1.318  & 1.289  & 1.419  \\
\hline
IGEs$~[M{\sun}]$        & 0.54  & 0.56   & 0.65   & 0.65   & 0.64   & 0.67   \\
  $[\%\, M_{tot}]$     & 38    & 39     & 40     & 37     & 35     & 33     \\
\hline
IMEs$~[M{\sun}]$        & 0.28  & 0.30   & 0.34   & 0.36   & 0.35   & 0.44   \\
  $[\%\, M_{tot}]$     & 20    & 20     & 20     & 21     & 20     & 22     \\
\hline
C+O$~[M{\sun}]$        & 0.58  & 0.61   & 0.65   & 0.73   & 0.80   & 0.90   \\
  $[\%\, M_{tot}]$     & 42    & 41     & 40     & 42     & 45     & 45      \\
\hline
\end{tabular}
\end{center}
\caption{Energetics and compositions for the deflagration model with $c3$~ignition in different rotators.}
\label{tab:sntd_rotlaw_c3}
\end{table}

\subsubsection{$c3$~ignition} \label{rotlaw_c3}

Fig. \ref{panel_etot_c3} shows the temporal evolution of the total
energy $E_{tot}$ for different rotators that are ignited with the 
$c3$~ignition. Note that the more massive WDs
start from more negative values of total energy. This is
caused by the larger gravitational energies $E_{grav}$ for heavier stars
(see Table \ref{tab:sntd_rotlaw_c3}). All simulations end in 
explosions that cause the star to unbind at $t \sim 1~\mathrm{s}$.
This takes longest for the heaviest WD, AWD3. The differences
in $E_{tot}$ that are apparent in the beginning become smaller during the explosion
process, which is caused by a higher amount of released nuclear energy
$E_{nuc}$ for the more massive WDs. 

The non-rotating star exhibits the highest value of
$E_{tot}$ in the homologous expansion phase.
$E_{tot}$ at $t = 5~\mathrm{s}$ consists mainly of kinetic energy $E_{kin}$ of the
ejecta for the typical
temporal evolution of the energy contributions. Accordingly, the
ejecta are fastest for the non-rotating progenitor on average.

With respect to the stellar composition after the explosion, the following
trend is visible from Table \ref{tab:sntd_rotlaw_c3} and
Fig. \ref{fig:composition_burn_rel}: 
the heavier the progenitor, the more fuel is left unburnt. The 
amount of IGEs after $5~\mathrm{s}$ ranges from $0.54~M_{\sun}$ for the
non-rotating star to $0.67~M_{\sun}$ for the rapidly rotating model
AWD3.
However, the fraction of IGEs with respect to the total mass
{\it decreases} for rapid rotation. Only $33~\%$ of the total
$2.02~M_{\sun}$ is converted to IGEs for the 
AWD3
rotator. The amount of
IMEs rises for more rapid rotation, its 
fraction is remarkably constant at $\sim 20~\%$ for all rotation
laws. Combined with the fact that the amount of unburnt material becomes larger
for faster rotators, this result indicates that the higher absolute amount of
IMEs for the rapid rotators is not obtained ``at the cost'' of fuel in
contrast to observations.
It is a consequence of the small amount of produced IGEs.

\subsubsection{Stochastic ignition} \label{rotlaw_stochignt}

Basically, the trends found in the case of the $c3$~ignition scenario remain
valid for explosions with stochastic ignition. However, the energetics of the
explosions show a larger variation (see Fig.~\ref{panel_etot_stoch} and 
Table~\ref{tab:sntd_rotlaw_stochignt}), and even less IGEs are produced 
while more unburnt fuel is left over (see Figs. \ref{fig:composition_burn_abs} and
\ref{fig:composition_burn_rel} as well as Table \ref{tab:sntd_rotlaw_stochignt}) . 
This is explained by the fact that,
unlike the symmetric $c3$~ignition, the stochastically
ignited bubbles are immediately subject to buoyant motion since the
net buoyancy is not balanced by a symmetric alignment of the ashes,
therefore having less time to burn the central region efficiently. 
Also fluctuations in the composition between different progenitor models
are more distinct for stochastic ignition compared to 
$c3$~ignition. This is related to the 
fact that not only the overall number of ignitions but also the
specific locations of the ignitions influence the
explosion outcome. 

The most energetic explosions with high fractions of IGEs result
from the non-rotating and the AWD1 
progenitor models. For the latter model, the central angular
velocity is low compared to the other differentially rotating models.
Since convection in the pre-supernova phases causes  
efficient angular momentum transport throughout the core,
however, the AWD1 rotator appears to be a rather unlikely
outcome of an evolutionary scenario leading to a thermonuclear
supernova.
For the ``AWD4 $C_e = 5 \times 10^{\,4}$\,''  explosion model, on the other hand, 
the amount of IMEs is somewhat higher, but this is compensated by little production of IGEs.
Consequently, a larger fraction of fuel remains unburnt. 
The ``$j_{const}$ $C_e = 5 \times 10^{\,3}$\,'' and
``$v_{const}$ $C_e =  5 \times 10^{\,3}$\,'' explosion models
generate yet smaller amounts of IGEs
(see Table \ref{tab:sntd_rotlaw_stochignt_vjconst}). However, neither the
$j_{const}$  nor the $v_{const}$ rotation law are likely to occur in
WDs. For the critically rotating rigid rotator, on the other hand,
approximately the same results are obtained as in the case of
the non-rotating star. Consequently, this type of rotating WD appears to be a
progenitor candidate for type Ia supernovae caused by pure deflagrations.

\subsection{Shear motion during the explosion} \label{slowdown}

It is a common belief that the shear introduced by differential
rotation can enhance the explosion strength by increasing the flame
surface. Although rotating WDs exhibit high rotation velocities (e.g., the
fastest regions of the AWD1 rotator move with $10~\%$ of the speed of light),
within the entire span of time during which burning takes place ($t
\lesssim 1.5~\mathrm{s}$) the direct influence of rotation on the flame
surface is limited. The 
rotators accomplish only about half a
rotation if the burning induced expansion is neglected.  
The latter
arranges a slowdown of the star as soon as 
the thermonuclear runaway is initiated as a result of angular momentum
conservation. 

Unless a distinct jump in $\Omega$ occurs --- which is not
the case for smooth differential rotation --- , the flame will not be
significantly influenced by rotation in a direct way (but, as
described in section \ref{impact_buoyancy}, by buoyancy and angular
momentum effects). 

\citet{2004A&A...419..623Y} proposed that a rotation could trigger a
deflagration-to-detonation transition 
since 
the outer layers still rotate rapidly when the outward burning flame causes
a slowdown of the WD. The resulting gradient in velocity due to burning is higher than
by differential rotation
and could possibly generate a detonation if the flame surface is
increased abruptly. 
In particular, such a rapid jump of angular velocity is found for the rotation
law of a WD merger \citep{YoonPodsiRoss}. 
We do not consider DDTs in this work; however, an investigation of shear-induced DDTs with the method introduced by \citet{GolombekNiemeyer2005} would be possible and interesting.
In particular, if observational evidence for rapid rotation of the progenitor star should be found, the DDT mechanism will be favored since prompt detonations of rapid rotators cannot explain normal SNe~Ia \citep{PAPER_II}.

\begin{table*}[t]%[tp]
\begin{center}
\begin{tabular}{|@{\,}r@{\,}||*{6}{@{\,}c@{\,}|}}
\hline
                         & norot & rigid / rigid$^*$ & AWD1 & AWD2 & AWD4 & AWD3 \\
\hline\hline
$M_{tot}~[M{\sun}]$          & 1.40  & 1.47           & 1.64   & 1.74   & 1.79   & 2.02   \\
\hline\hline
$C_e$               & $5\times10^{\,3}$ & $5\times10^{\,3}$ & $5\times10^{\,3}$ & $5\times10^{\,3}$ & $5\times10^{\,4}$ & $5\times10^{\,3}$  \\
\hline
$\Omega~[\pi]$              &   2   &   2            &    2   &   2    &   2    &   2    \\   
\hline\hline
$I_{2\pi}$                  & 172   & 167    /  163   & 168    & 154    & 404 & 152   \\
\hline\hline
\multicolumn{7}{|l|} {t~=~0\,s} \\
\hline
$E_{grav}~$[10$^{\,51}$\,erg] &-3.027 & -3.204          & -3.665 & -4.027 & -4.135 & -4.518  \\
\hline
$E_{rot}~$[10$^{\,50}$\,erg]  & 0.00  & 0.44            & 1.75   & 2.62   & 2.99   & 4.46   \\
\hline
$\beta~[\%]$           & 0.00   & 1.38            & 4.78   & 6.50   & 7.22   & 9.87   \\
\hline\hline
\multicolumn{7}{|l|} {t~=~5\,s} \\
\hline
$E_{kin}~$[10$^{\,50}$\,erg]  & 8.69  & 7.70 / 5.63     & 8.64  & 3.67   & 3.68  & 3.61   \\
\hline
$E_{tot}~$[10$^{\,50}$\,erg]  & 8.16  & 7.04   /  4.82  & 7.84   & 1.95   & 1.97  & 1.33   \\
\hline
$E_{nuc}~$[10$^{\,51}$\,erg]  & 1.282 & 1.234  / 1.005  & 1.495  & 1.016  & 1.073 & 1.182  \\
\hline
IGEs~$[M{\sun}]$         & 0.67  & 0.63 / 0.50     & 0.79   & 0.47   & 0.52   & 0.51   \\
  $[\%\, M_{tot}]$      & 48    & 43 / 34         & 48     & 27     & 29     & 25   \\
\hline
IMEs~$[M{\sun}]$         & 0.28  & 0.30 / 0.27     & 0.32   & 0.34   & 0.31   & 0.46   \\
  $[\%\, M_{tot}]$      & 20    & 20 / 19         & 20     & 20     & 17     & 23   \\
\hline
C+O~$[M{\sun}]$         & 0.45  & 0.54 / 0.70     & 0.53   & 0.93   & 0.96   & 1.04   \\
  $[\%\, M_{tot}]$      & 32    & 37 / 47         & 32     & 53     & 54     & 52   \\
\hline
\end{tabular}
\end{center}
\caption{Energetics and compositions for the deflagration model 
  activated by the stochastic ignition with $C_e = 5\times10^{\,3}$
  ($C_e = 5\times10^{\,4}$ for AWD4) in different rotators.}
\label{tab:sntd_rotlaw_stochignt}
\end{table*}

\begin{table}[t]%[tp]
\begin{center}
\begin{tabular}{|r||*{6}{c|}}
\hline
                            & $j_{const}$ & $v_{const}$ \\
\hline\hline
$M_{tot}~[M{\sun}]$         & 1.80               & 1.71        \\
\hline\hline
$C_e$                       & $5\times10^{\,3}$  & $5\times10^{\,3}$ \\
\hline
$\Omega~[\pi]$              &   2                &   2       \\   
\hline\hline
$I_{2\pi}$                  & 159                & 158      \\
\hline\hline
\multicolumn{3}{|l|} {t~=~0\,s} \\
\hline
$E_{grav}~$[10$^{\,51}$\,erg] & -4.271           & -4.056       \\
\hline
$E_{rot}~$[10$^{\,50}$\,erg]  & 3.16             & 2.39          \\
\hline
$\beta~[\%]$                  & 7.40             & 5.89        \\
\hline\hline
\multicolumn{3}{|l|} {t~=~5\,s} \\
\hline
$E_{kin}~$[10$^{\,50}$\,erg]  & 1.66             & 1.29    \\
\hline
$E_{tot}~$[10$^{\,50}$\,erg]  & -0.75             & -1.11     \\
\hline
$E_{nuc}~$[10$^{\,51}$\,erg]  & 0.817            & 0.679    \\
\hline
IGEs~$[M{\sun}]$             & 0.34               & 0.29   \\
  $[\%\, M_{tot}]$          & 19                 & 17      \\
\hline
IMEs~$[M{\sun}]$             & 0.34               & 0.27    \\
  $[\%\, M_{tot}]$          & 19                 & 16     \\
\hline
C+O~$[M{\sun}]$             & 1.12               & 1.15    \\
  $[\%\, M_{tot}]$          & 62                 & 67       \\
\hline
\end{tabular}
\end{center}
\caption{Energetics and compositions for the deflagration model
  activated by stochastic ignition with $C_e = 5\times10^{\,3}$
  in the $j_{const}$ and $v_{const}$ rotators.}
\label{tab:sntd_rotlaw_stochignt_vjconst}
\end{table}
\begin{figure*}[th]
  \mbox{
  \subfigure[``norot $c3$''explosion model]
  {\includegraphics[width=8.5cm]{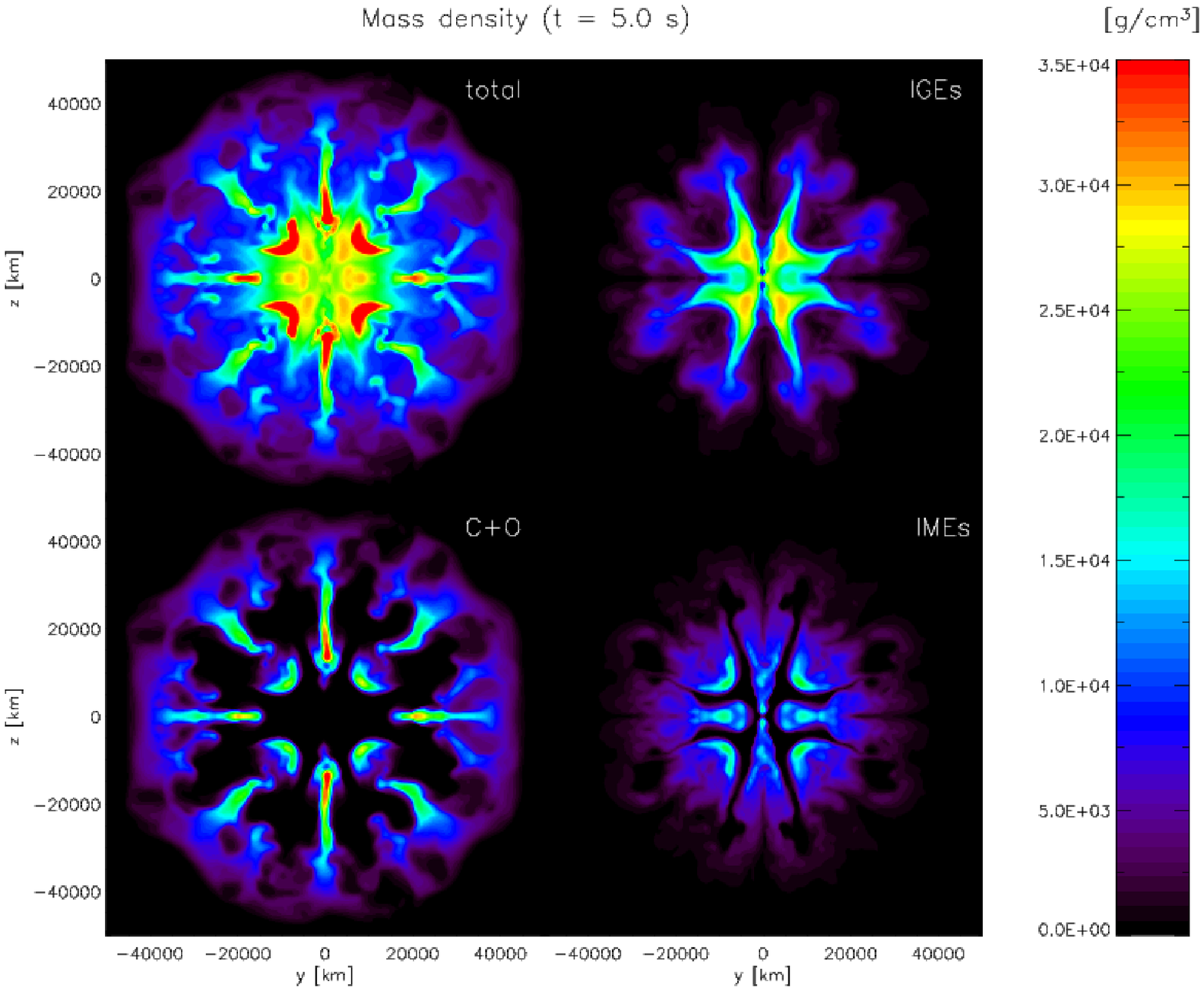}
    \label{fig:norot_frac_c3}}
  \subfigure[``AWD3 $c3$'' explosion model]
  {\includegraphics[width=8.5cm]{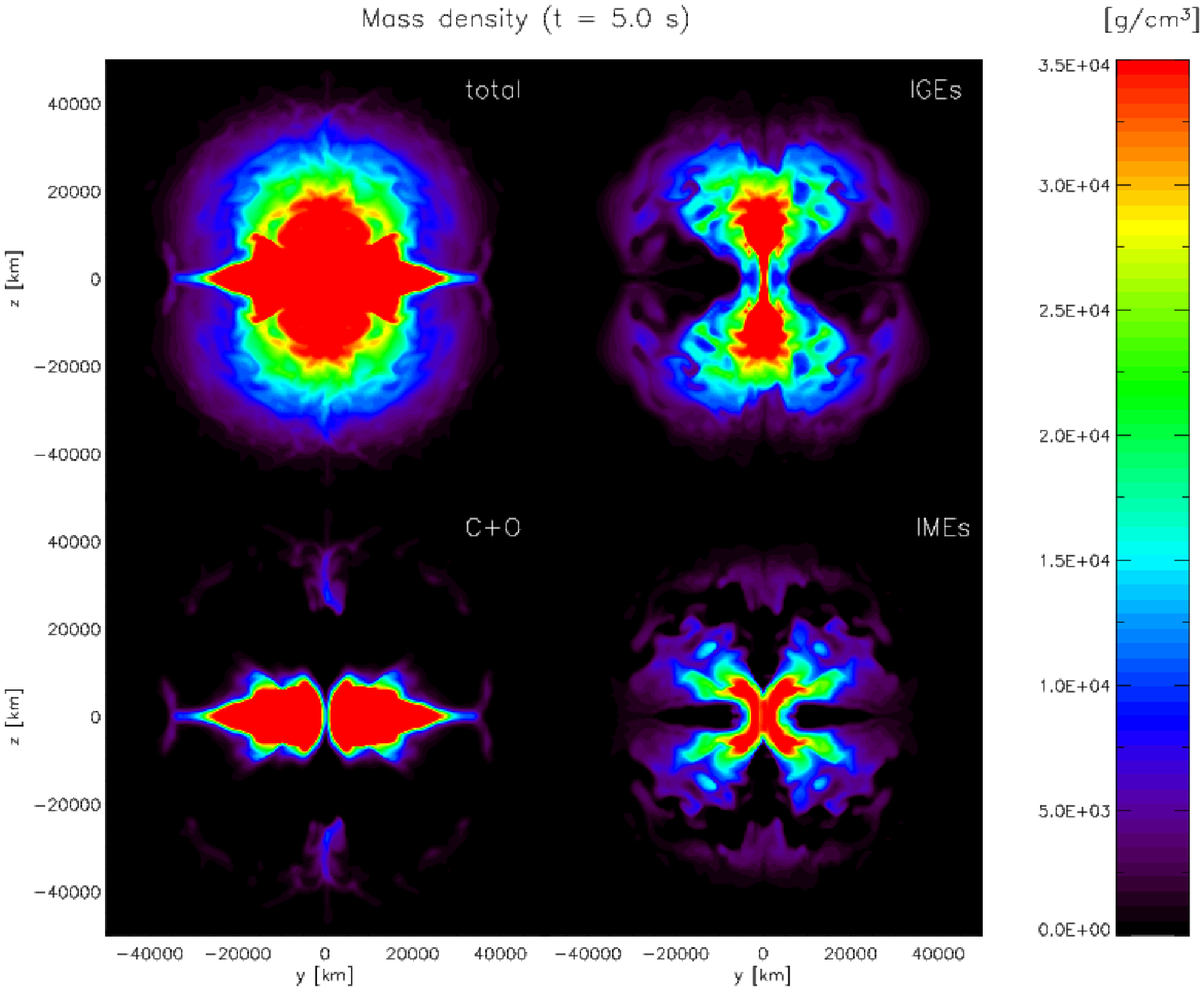}
    \label{fig:20AWD3_frac_c3}}}
  \caption{Total (upper left) and fractional (upper right: IGEs; lower left:
  unburnt fuel; lower right: IMEs) mass densities after $t =
  5~\mathrm{s}$ for the ``norot $c3$''
  (a) and ``AWD3 $c3$''
  (b) explosion models. A mixing of the
  species throughout the 
  star emerges in both scenarios, particularly the rapid rotator leaves
  fuel in the stellar core. Cross sections along the rotation axis of simulations exhibiting
  equatorial symmetry are shown.} 
  \label{fig:20AWD3_frac_both}
\end{figure*}

\subsection{Expected spectral features} \label{spec_highv}

An inspection of the composition and the kinematics of the ejecta in
the homologous expansion phase allows a prediction of spectral
features even without employing a detailed postprocessing
study. Regarding rotation, the most serious discrepancy between the
simulation results and observations is the prediction of C and O at
low radial expansion velocities, i.e., close to the centre. This
result, already a problem for the non-rotating deflagration model,
grows more acute for rapid rotation.  
Fig. \ref{fig:20AWD3_frac_both} shows the total and fractional mass
densities after $t = 5~\mathrm{s}$ for the ``norot
$c3$'' and ``AWD3 $c3$'' explosion models. For the sake of demonstration, only the
$c3$~ignition is presented here. The results are qualitatively in
accordance with those of the stochastic ignition. 

\begin{figure}[th]
  \begin{minipage}{0.5\textwidth}
    \centering
    \subfigure[``norot $C_e = 5 \times 10^{\,4}$\,'']
    {\includegraphics[width=\linewidth]{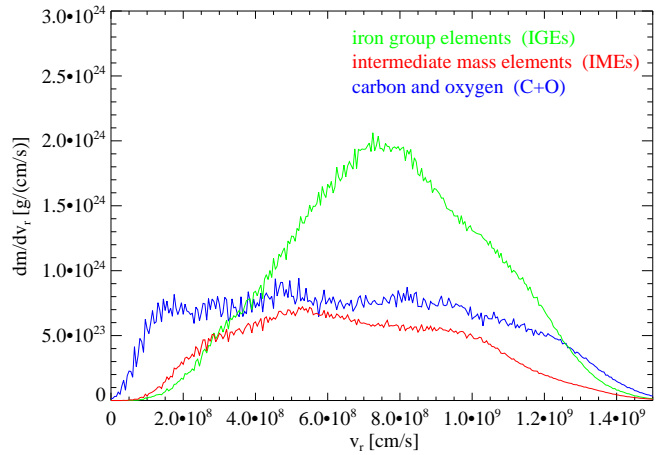}
      \label{fig:norot_massdist_velspace}}
 \end{minipage}\hfill
  \begin{minipage}{0.5\textwidth}
    \centering
    \subfigure[``AWD4 $C_e = 5 \times 10^{\,4}$\,'']
    {\includegraphics[width=\linewidth]{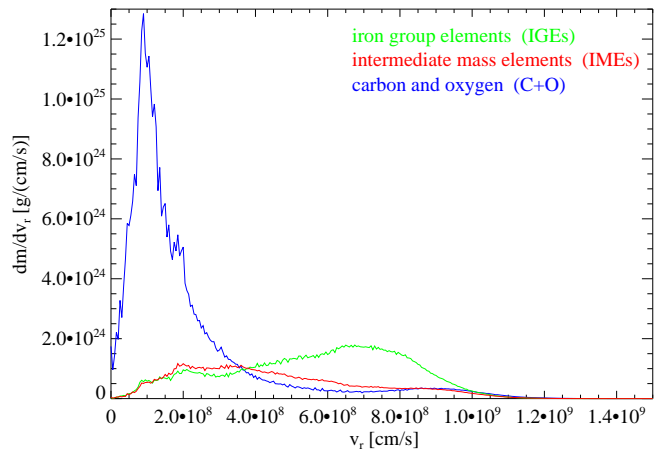}
      \label{fig:14AWD4_massdist_velspace}}
  \end{minipage}
  \caption{Probability density functions in radial velocity space for
  the ``norot $C_e = 5 \times 10^{\,4}$\,''
  (a) and ``AWD4 $C_e = 5
  \times 10^{\,4}$\,'' (b) explosion models}
  \label{fig:both_massdist_velspace}
\end{figure}

SN Ia observations indicate that the IMEs should be located in
the outer regions and enclose the IGEs for normal SNe~Ia,
whereas C and O must not be present near the centre at all. However,
as a result of the turbulent combustion whose driving force is the
Rayleigh-Taylor instability, the species are mixed throughout
the stellar interior. Furthermore, the existence of a significant amount of
unburnt fuel in the core is inconsistent with observations. 
The following trend that becomes stronger for more rapid
rotation emerges:
Initially, burning mostly advances towards the stellar poles and
generates IGEs. Later, when the flame spreads toward
the equatorial plane and the density has already dropped, IMEs are
produced. However, the burning process ceases before the 
central part and material close to the equatorial plane can be burnt. 
This is reflected by the distinct peak of C and O at low radial
velocities that can be seen in Fig.~\ref{fig:both_massdist_velspace} for the 
AWD4 rotator. For comparison, the distributions in radial velocity space
are also plotted for the non-rotating stochastic ignition model.

\begin{figure}[th]
\centering
\includegraphics[width=0.95\linewidth,clip]{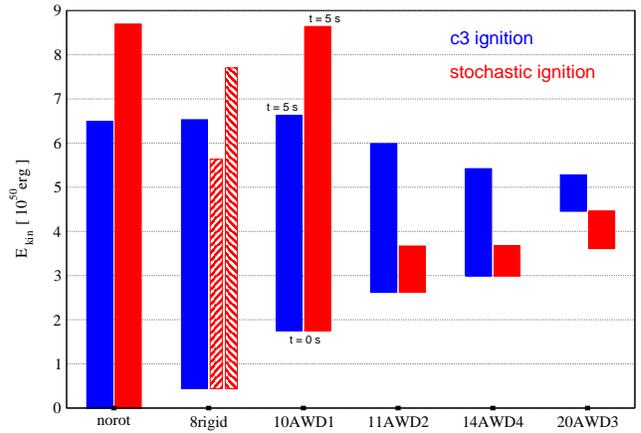}
\caption{Initially available
rotational energy (bottom of the bars) and kinetic energy of the
ejecta after $t = 5~\mathrm{s}$ (top of the bars) for successively heavier WD models}
\label{fig:high_v_integral}
\end{figure}

The highest radial velocities in the non-rotating case are 
$\lesssim 15 \times 10^{\,8}~\mathrm{cm/s}$. 
Consequently, high velocity features cannot be
explained by the deflagration model, although higher velocities
might become apparent with increasing resolution.
In the rotating case, the radial velocities are even smaller 
($\lesssim 10 \times 10^{\,8}~\mathrm{cm/s}$),
despite the fact that rotating progenitors include
a large amount of rotational kinetic
energy. This is due to the greater gravitational
attraction of the rotating stars 
(cf. Table \ref{tab:sntd_stochignt_20AWD3}), and in addition the released
energy is comparatively small because of incomplete burning.
Fig. \ref{fig:high_v_integral} compares the initially available
rotational energy (bottom of the bars) and the kinetic energies of the
ejecta after $t = 5~\mathrm{s}$ (top of the bars) for different
WD models. Except for the AWD1 rotator without
convective core, the disposable 
kinetic energies become successively smaller for increasing 
strength of rotation. Note that for the 
``AWD3 $C_e = 5 \times 10^{\,3}$\,'' explosion model, the kinetic
energy after $t = 
5~\mathrm{s}$ is even less than the kinetic energy initially present
in the rotation motion.

\section{Conclusions}
\label{concl}

In this paper, we investigated the influence of nearly critical rotation of the
progenitor star on SN~Ia explosions. We performed three dimensional numerical
simulations of thermonuclear deflagrations 
for different rotation laws of the white dwarf progenitors and various
initial conditions. The influence of the ignition process turns out to be larger in
rotating stars compared to their non-rotating
counterparts, which might contribute to the observed diversity among SNe~Ia.

The main result of this study is that the amount of iron group elements 
is not significantly increased compared to the non-rotating case, although the
rotating WD models are 
notably heavier than their non-rotating counterparts. Due to the
centrifugal expansion, rotators contain more material at low densities in
general, thus the amount of material capable of burning to iron group elements
is only weakly increased if the star is burnt by a deflagration. 
Furthermore, due to rotationally 
induced anisotropic buoyancy effects and, at the same time, inhibition of
effective mixing parallel to the equatorial plane,
the flame preferentially propagates
toward the stellar poles. This leads to comparably weak 
and anisotropic explosions 
that leave behind unburnt material at the centre and in the
equatorial plane. Moreover, no significant effects of shear
motion acting on the flame are observed. Deflagrations of rapid
rotators do not
cause high velocity features but result in overall low expansion
velocities.  
As a consequence, the pure deflagration scenario is ruled out
in the case of rapid rotation of the progenitor star. The incineration of
the critical rigid rotator shows similar features as the non-rotating
scenario. Therefore, deflagrations could be possible for
critical rigid rotation.

In conclusion, rotation of the progenitor star is unlikely to be the parameter
that causes the observed variation in peak luminosities among SNe~Ia.
Rapid rotation of the progenitor
star  derived by an accretion study can lead
to a great variety in explosion strengths in the deflagration scenario,
but only within a range that is ruled out for observational reasons. 
Otherwise, even the possible critical rigid rotation can not account for a
significant spread in the explosion outcome.

\begin{acknowledgements} 

We thank Ewald M\"uller, Friedrich R\"opke and Sung-Chul Yoon for
helpful comments and discussions.

\end{acknowledgements}

\bibliographystyle{aa}
\bibliography{12032_colour}

\end{document}